\documentstyle[12pt,psfig]{article}

\def\reference{\parskip 0pt\par\noindent\hangindent 0.5 truecm}

\begin{document}

\title{The First Results from the Parkes Multibeam High-Velocity Cloud Survey}

\author{Mary E. Putman $^{1}$ \and
 Brad K. Gibson $^{2}$ 
} 

\date{}
\maketitle

{\center
$^1$ Mount Stromlo \& Siding Spring Observatories, Australian National
University, Weston Creek P.~O., Weston, ACT, 2611  Australia
\\putman@mso.anu.edu.au\\[3mm]
$^2$ Center for Astrophysics and Space Astronomy, University of Colorado, 
Campus Box 389, Boulder, Colorado, 80309  USA
\\bgibson@casa.colorado.edu\\[3mm]
}

\begin{abstract} 

With the onset of the HI Parkes All-Sky Survey, a new view of the
high-velocity cloud (HVC) distribution in the southern sky is being revealed.
The dense spatial sampling and unbiased coverage of HIPASS gives it 
multiple advantages over previous surveys of the southern sky.  Detailed views of
the clouds' structure and large mosaics of the HI sky allow us to link individual 
concentrations to larger structures (such as the Magellanic Clouds and the Galaxy), 
providing pictorial clues as to the origins of HVCs. 
 It is clear that HVCs cannot be uniformly assigned a single origin scenario, and must be
categorised appropriately. 

\end{abstract}


\bigskip

\section{Introduction}

High-velocity clouds (HVCs) are concentrations of neutral hydrogen which
have velocities forbidden by simple models of Galactic rotation.
Their velocities, with respect to the Local Standard of Rest,
typically fall in the range of $|v_{lsr}|\approx 80-400$ km 
s$^{-1}$,
depending on the direction of observation.\footnote{There is a well-known
north-south hemisphere asymmetry in the HVC velocity distribution, 
such that the maximum $|v_{lsr}|$ seen in
the north is $\sim 450$ km s$^{-1}$, $\sim 100$ km s$^{-1}$ greater than the
maximum seen in the south.}
Despite covering $\sim$ 37\% of the sky (Murphy, Lockman \& Savage 1995),
a consensus has yet to be reached on their origin.  
Several models have been proposed, both Galactic
(e.g. supernova shells, warped structure in the
disk, Galactic fountain infall/outflow) and extragalactic (e.g. remnants of the
formation of Local group, tidally-disrupted material from the Magellanic Clouds
or other dwarf galaxy companions), but theory alone has not been able to
discriminate between these scenarios (and as we discuss here, ascribing
the entire HVC population to one in particular would probably be a mistake).  Wakker \& van
Woerden (1997) provide a timely review of the subject, both from an
observational and theoretical point of view.  In principle,
observational tests can be devised which can discriminate (roughly) between
Galactic and extragalactic models, but to date, the implementation of such
tests (e.g. direct distance determinations, metallicity, or emission measure
properties) has proven difficult.  Each of these observational tests depends
on an understanding of the spatial and kinematical structure of HVCs, and 
while such high-quality HI
data does exist in the north (e.g. Hulsbosch \& Wakker 1988; Hartmann \& Burton
1997), the southern sky has not enjoyed the same status.

To date the most complete investigation in the south is due to
Bajaja et~al. (1985, hereafter B85), which sampled the sky on a 
$2^{\circ} \times 2^{\circ}$ grid with the IAR 30m 
at Villa Elisa.  Next year should see the release of the
updated Villa Elisa Southern Sky Survey (VESSS, Morras et~al. 1999), 
the southern analog to the
Leiden-Dwingeloo Survey (LDS, Hartmann \& Burton 1997), but for now one is
restricted to the B85 survey for southern HVCs (and indeed, this was the data
source for the southern half of the canonical catalog of HVCs by
Wakker \& van Woerden 1991).  The B85 spatial sampling and beam size is wholly 
unsatisfactory for unveiling the presence of fine-scale structure and any 
large population of Compact HVCs (henceforth, CHVCs). 
In the subsequent sections, we present results from the first complete,
fully sampled, southern sky survey for HVCs. 
Our initial results reveal intricate detail in the spatial structure of
southern HVCs, a large population of undiscovered CHVCs, and 
filamentary connections between previously assumed discrete clumps. 

\section{Observations \& Data Reduction}

The HI Parkes All-Sky Survey (HIPASS) is a blind HI survey
of the southern sky ($\delta\le 0^\circ$), over the velocity range
$-1,200$ to 12,700 km s$^{-1}$ (Staveley-Smith 1997).  It therefore encompasses the 
Milky Way, the Magellanic Clouds, and the more distant
 Universe.  The survey is conducted with the Parkes 64m radio telescope
equipped with a multibeam receiver, a focal-plane array of 13
feeds set in an hexagonal grid.  The
spectrometer has 1024 channels for each polarisation and beam,
with a channel spacing of 13.2 km s$^{-1}$. 
The survey is conducted by actively
scanning the telescope in 8$^{\circ}$ 
strips of declination, and displacing adjacent scans in RA such
that sub-Nyquist sampling is achieved. The complete
survey will scan the sky five times, with an effective integration time of
$\sim 5$ minutes per sky point, corresponding to an rms 
brightness temperature sensitivity of $\sim 9$ mK.  The data presented in this
paper only incorporates the first set of scans (which still Nyquist
sample the sky), corresponding to an rms
sensitivity of  $\sim 20$ mK.  

The data were reduced using a modified
version of the standard {\tt aips${++}$} based HIPASS reduction software (which was
designed for imaging discrete HI sources) (Barnes 1998). 
To recover the extended structure of HVCs, we used
the entire 8$^{\circ}$ scan to calculate the bandpass correction for a 
particular channel and polarisation.  
Several statistical methods were tested, with the chosen method a compromise 
between recovering larger structures and the increasing variation in the 
correction for adjacent scans.  
The method chosen breaks
each 8$^{\circ}$ scan into five sections, finds the median flux in each section
and uses the minimum of the five medians for the bandpass correction.
This greatly increases our ability to recover large scale structure, except
in areas where the emission fills more than 90\% of a given 
scan (e.g. near the Galactic Plane).  To minimise the effect of the
ripple from Galactic HI, the data is Hanning smoothed, resulting in a 
velocity resolution of 26.4 km s$^{-1}$.
Each individual scan was gridded using a combination of {\tt aips${++}$}, {\tt aips} and
{\tt miriad}.  The gridding process, and active scanning method used, introduce
some striping into the final images.  The gridding also increases the FWHP beamwidth from $14^{\prime}.4$ to an
effective value of $\sim15^\prime.5$.

\section{The Multibeam Advantage}

\begin{table}
\caption{HIPASS survey parameters versus those of B85 and VESSS.}
\begin{center}\normalsize
\begin{tabular} {cccc}\hline
Parameter & HIPASS & B85 & VESSS \\\hline
Telescope & Parkes 64m  & IAR 30m & IAR 30m\\
Sky & $\delta<0^\circ$ & $\delta<-15^\circ$ & $\delta<-25^\circ$ \\
HPBW & $15^{\prime}.5$ & $34^{\prime}$ & $34^{\prime}$\\
Grid & $\sim6^{\prime}$ & $120^{\prime}$ & $30^{\prime}$\\
$\Delta$v (km s$^{-1}$) & 26.4 & 16 & 8\\
T$_{B}$(K) ($5\sigma$) & 0.045 (5 scans) & 0.125 & $\sim 0.1$\\
 &  0.1 (1 scan) & &\\
N$_{HI}$(cm$^{-2}$) ($5\sigma$) & 1.6 $\times 10^{18}$ (5 scans) & 
$4.6 \times 10^{18}$ & \\
($\Delta$v = 20 km s$^{-1}$) & 3.6 $\times 10^{18}$ (1 scan) & & \\\hline
\end{tabular}
\end{center}
\end{table}

The key parameters of HIPASS are contrasted with the B85 and VESSS southern
surveys in Table 1.  The primary advantages of HIPASS are its
superior spatial resolution and dense spatial sampling.  If HIPASS's active
scanning method was put on a point-and-shoot grid it would correspond to
a grid spacing of $\sim 6^\prime$ with a $\sim 15^\prime.5$ beam.
The HIPASS survey zone also covers almost twice as much of the sky as the Argentinian
surveys, providing desired consistency checks 
with the northern hemisphere HVC surveys.  
Some of the differences alluded to in Table 1 are best expressed in a picture.
Figure 1 shows the same region of sky as viewed by HIPASS and B85.  Clearly,
the gross properties are visible in the B85 data, but just as clear is the
startling improvement provided by the HIPASS data.  The low resolution and
sparse sampling of B85 tends to merge many individual features into a single,
amorphous blob; the complex structure of the clouds only becomes apparent when 
examining the HIPASS data.

We are already in a position to speculate on the expected increase in the
number of catalogued, isolated, HVCs and CHVCs.  Based on a visual inspection
of the data in hand, we expect this number to increase by a factor of
five.  This is not to say that the integrated HI flux or mass will
increase by a factor of 5, but that the \it number \rm (especially of CHVCs)
is expected to increase dramatically.  These results will greatly affect HVC catalog
statistics.  What previously appeared to be a lack of positive velocity clouds, may
simply have been the inability of previous surveys to resolve the individual
clumps.  Figure 2 offers further evidence to this
effect; the boxes therein
mark the positions of \it all \rm
previously catalogued clouds in this area (from Wakker \& van Woerden 1991).  
Again, the large population of individual HVCs missed by B85 (and thus the Wakker \&
van Woerden compilation) is clearly evident.   We will return to Figure 2 shortly.

Just as important as the discovery of enormous numbers of new clouds,
is the parallel
increase in positional accuracy of existing (and new) HVCs.  
This has important implications for
optical and UV follow-up programs to determine HVC distances and
metallicities.  Both programs hinge upon knowledge of an HVC's intervening HI column
density along the line of sight to a suitable stellar or extragalactic
probe.  Basing these programs upon the older, low resolution data is risky, but was 
in the past, a necessity.  As alluded to in Figure 1, HIPASS will improve this situation 
dramatically.  Looking at the B85 data, the background Seyfert galaxy ESO~265-G23 was 
thought to be aligned with the
extreme positive HVC shown; the HIPASS data shows that it is actually situated in what appears to be an
HI-free region.  Ultimately, even higher-resolution synthesis
observations are needed (a program we are
pursuing at the ATCA), but obviously HIPASS offers a
superior base to build upon when prioritising candidate background probes.
Similar arguments can be made when drawing up target lists for follow-up
emission-line studies, a program we are pursuing with J. Bland-Hawthorn at the
AAT and WHT (see Bland-Hawthorn et~al. 1998).

Another advantageous feature of HIPASS is its unbiased sky coverage.  The displayed population of
positive HVCs in Figure 2 look fairly unremarkable when framed within a $\sim 300$\,deg$^2$ 
region.  It is only upon expanding one's view
by an order of magnitude that their true nature can be appreciated.  The sheer
size of the HIPASS dataset makes such ``big picture'' views easily attainable.
Figure 3 takes the clouds of Figure 2, and places them into the context of a 
2400 deg$^2$ area of sky.  Only now does this
population of HVCs' relationship to the Magellanic System become apparent.
Their link to the Magellanic Clouds becomes even more evident in the individual channel
maps.  Figure 4 shows the channel centred upon $v_{\rm lsr} = 323$ 
km s$^{-1}$ and depicts
the feature's spatial continuity for over $25^{\circ}$, starting from a 
region in the Magellanic Bridge.

This continuous feature on the leading side of the Magellanic Clouds, highlighted
by Figure 4, is a natural prediction of tidal models which simulate the
dynamical interaction of the Clouds with the Milky Way (e.g. Gardiner \&
Noguchi 1996).   Our discovery of this Leading Arm was reported in Putman 
et~al. (1998); parts of the feature are apparent in the old Mathewson
\& Ford (1984) data, but it was only the advent of the HIPASS reductions which
allowed us to demonstrate the spatial and kinematic continuity of this feature. The lack of
continuity had been used by proponents of ram pressure Magellanic Stream
formation models to claim the tidal model was
excluded (e.g. Moore \& Davis 1994); on the contrary, our finding of a continuous 
Leading Arm which emanates from the Magellanic System has allowed us to eliminate virtually all pure ram pressure
models.

\section{HVC Structure}

Continuous streamer structure is a common phenomenon which 
appears in the fully sampled HIPASS data.   
The Magellanic Stream (which trails for $\sim$100$^{\circ}$ behind the Clouds in Figure
3) and Leading Arm, discussed in the previous section, are obvious examples.  
Another example of a spatially continuous filament was shown in Figure 1; this
feature also has a strikingly systematic velocity structure (see Putman \& Gibson 1998 or
Morras \& Bajaja 1983).  
Figure 1 (and also Figure 4) can be put into further
context by examining Figure 5.  Here, the filamentary structure of Figure 1
(shown at the top of Figure 5) appears to be an extension of the Leading Arm shown
in Figure 4, the entire structure
perhaps representing a pseudo-continuous, $\sim 50^{\circ}$-long, leading
tidal feature.  Such a scenario is supported by the metallicity determination
for the dominant filament in Figure 1 from Lu
et~al. (1998), which matches that of the Magellanic Clouds.  The ``kink'' in
the Leading Arm above the Galactic Plane, though, appears to be
difficult to recover under a tidal scenario, as well tidal
models which incorporate a small degree of drag (Gardiner 1999).

We should stress that while filamentary HI structure may be
common, assigning a tidal origin to all such features would certainly be
incorrect.  There are many situations where the filamentary HVC 
actually connects to Galactic emission when it
is traced back in position \it and \rm velocity.  This is demonstrated in Figure 6, where the knot 
of emission in the $v_{lsr}=86$ km s$^{-1}$ channel
corresponds to HVC\#350 in the Wakker \& van Woerden (1991) compilation.  While
technically this is an HVC because of its anomalous velocity, Figure 6 makes it
readily apparent that it is not
an isolated HVC, but an extension of Galactic emission. 
This example reiterates an earlier point, that clearly not all HVCs are
created by the same mechanism - while HVC\#350 appears to be associated with
the Galaxy, those of Figure 5 are almost certainly (predominantly) associated
with tidal stripping of material from the Magellanic System.  
Figure 6 demonstrates that blind adoption of velocity cut-offs in any HVC
catalog construction is a dangerous practice which does not necessarily
take into account the ``big picture''.

It is tempting to associate stream-like ``HVCs'', such as those shown in Figure
6, with the continuation of
the worms defined by Koo, Heiles \& Reach (1992), possible supernova 
break-outs, or the splash-down effects
of discrete HVCs.  
However, as noted already, 
continuous streams of anomalous HI, such as the Magellanic Stream and
its associated Leading Arm, which
are clearly separate from Galactic emission, almost certainly reflect the tidal
disruption of our neighbours (both present-day \it and \rm past).  Perhaps the
overall filamentary structure of the clouds is also
related to magnetic fields which extend into the Galactic halo (Kalberla \&
Kerp 1998)?  Dense clumps of emission, or CHVCs in the nomenclature of
Braun \& Burton (1999),\footnote{Braun \& Burton (1999) define a CHVC as an isolated cloud which 
has an angular size of less 
than $2^{\circ}$ FWHM.} may represent a distinct category, populating the
intergalactic medium of the Local Group, as suggested by Blitz et~al. (1999).
On the other hand, they often appear to be associated in position and velocity 
with neighbouring large-scale complexes, and they could simply be small bits 
of interactive debris, as demonstrated in the hydrodynamical
simulations of Li \& Thronson (1998).  Mapping the H$\alpha$ emissivities of
these CHVCs, along with their neighbouring complexes, may provide clues as to
any differences which might exist in the ionising environments in
which they reside.

The complex structure of HVCs defines our need for dense spatial sampling
and high resolution observations.   As shown here, the Parkes Multibeam 
HVC Survey has the 
capability to greatly increase our understanding of HVC distribution and
structure, with the Leading Arm feature a perfect example of its success
in uncovering new features in the Galactic halo.  Categorising clouds
according to their spatial and velocity distribution will be a crucial
future step for uncovering the origin of various high velocity complexes.
We are actively pursuing a program of H$\alpha$ emission studies,
high-resolution 21cm synthesis work, and FUSE (Far Ultraviolet Spectroscopic
Explorer) and GHRS (Goddard High-Resolution Spectrograph) metallicity
determinations for a number of HVCs which populate a wide variety of
environments.
By putting these results into the context of the Parkes Multibeam
HVC Survey we should be able to provide key clues to HVCs' mysterious origin(s).


\section*{Acknowledgements}

We thank the Multibeam Survey Working Group for their help in the HIPASS data 
collection, and the development of the instrumentation and software.  The
assistance of Bart Wakker, in preparing part of Figure 1, is likewise acknowledged.

\section*{References}

\reference Bajaja, E., et al. 1985, ApJS, 58, 143
\reference Barnes, D.G., Staveley-Smith, L., Ye, T. \& 
Oosterloo, T. 1998, in Astronomical 
Data Analysis Software and Systems (ADASS) VII, eds. Albrecht, R., Hook, 
R.N. \& Bushouse, H.A., ASP Conf. Series, in press
\reference Bland-Hawthorn, J., Veilleux, S., Cecil, G.N., Putman, M.E., Gibson, B.K.
\& Maloney, P.R. 1998, MNRAS, 299, 611
\reference Blitz, L., Spergel, D.N., Teuben, P.J., Hartmann, D., Burton, W.B. 1998, ApJ
submitted
\reference Braun, R. \& Burton, W.B. 1999, A\&A, in press
\reference Burton, W.B. 1988, in Galactic \& Extragalactic Radio Astronomy, 
eds. Verschuur, G.L. \& Kellermann, K.I., Springer-Verlag, p. 295
\reference Gardiner, L.T. in Stromlo Workshop on High-Velocity
Clouds, eds. Gibson, B.K. \& Putman, M.E., ASP Conf. Series, in press
\reference Gardiner, L.T. \& Noguchi, M. 1996, MNRAS, 278, 191
\reference Hartmann, D. \& Burton, W.B. 1997, Atlas of Galactic Neutral
Hydrogen, Cambridge Univ. Press
\reference Hulsbosch, A.N.M. \& Wakker, B.P. 1988, A\&AS, 75, 191
\reference Kalberla, P.M.W. \& Kerp, J. 1998, A\&A, 339, 745
\reference Koo, B., Heiles, C. \& Reach, W.T. 1992, ApJ, 390, 108
\reference Li, P.S. \& Thronson, H.A., 1999, in IAU Symp. 190 New Views of the 
Magellanic Clouds, eds. Chu, Y.-H., et~al., ASP Conf. Series, in press
\reference Lu, L., et al 1998, AJ, 115, 162
\reference Mathewson, D.S., Cleary, M.N. \& Murray, J.D. 1974, ApJ, 190, 291
\reference Mathewson, D.S. \& Ford, V.L. 1984, in Structure and Evolution of
the Magellanic Clouds, eds. van den Bergh, S. \& de Boer, K.S., Kluwer,
Dordrecht, p. 125
\reference Moore, B. \& Davis, M., 1994, MNRAS, 271, 209
\reference Morras, R., et~al. 1999, in Stromlo Workshop on High-Velocity 
Clouds, eds. Gibson, B.K. \& Putman, M.E., ASP Conf. Series, in press
\reference Morras, R. \& Bajaja, E. 1983, A\&AS, 51 131
\reference Murphy, E.M., Lockman, F.J. \& Savage, B.D. 1995, ApJ, 447, 642
\reference Putman, M.E. \& Gibson B.K. in Stromlo Workshop on High-Velocity
Clouds, eds. Gibson, B.K. \& Putman, M.E., ASP Conf. Series, in press
\reference Putman, M.E., Gibson, B.K., Staveley-Smith, L. et al. 1998, 
Nature, 394, 752
\reference Staveley-Smith, L. 1997, PASA, 14, 111
\reference Wakker, B.P. 1991, A\&A, 250, 499
\reference Wakker, B.P. \& van Woerden, H. 1991, A\&A, 250, 509
\reference Wakker, B.P. \& van Woerden, H. 1997, ARA\&A, 35, 217

\newpage

\begin{figure}
\caption{Column density maps of HIPASS data (top plot) and data from the 
Bajaja et al (1985) survey (bottom plot).  The maps cover the same $\sim
1000$\,deg$^2$ region of sky, centred upon the Extreme Positive Population of
HVCs (adopting the nomenclature of Wakker \& van Woerden 1991).  The velocity
range included here is $v_{lsr}=170 - 400$ km s$^{-1}$.  
The star symbol represents the position
of the Seyfert 1, ESO~265-G23 (see text).}
\end{figure}

\begin{figure}
\caption{Peak intensity map of positive velocity HVCs ($v_{lsr}=85$ - 400 km s$^{-1}$).  The solid boxes
represent the positions of \it all \rm (i.e., 7) 
clouds  in the Wakker \& van Woerden (1991) compilation.}
\end{figure}

\begin{figure}
\caption{Peak intensity map of a 2400 deg$^2$ region centred upon
the South Celestial Pole, including the region shown in Figure 2.  The velocity
range encompassed here is $v_{lsr}=85 - 400$ km s$^{-1}$.
The Large and Small Magellanic Clouds, the Magellanic Bridge and the 
beginning of the Magellanic Stream are shown at the top of the figure.  
The Galactic Plane, which
extends out to 120 km s$^{-1}$ in the direction $(\ell,b)=(300^\circ,0^\circ)$ (Burton 1988), is seen at 
the bottom of the figure.}
\end{figure}

\begin{figure}
\caption{A detailed view of the Leading Arm at the channel centred upon
$v_{\rm lsr} = 323$ km s$^{-1}$.
Contours are from 10 - 90\% of the brightness temperature maximum 
(T$_{\rm B}=0.88$ K).   The link
between the Magellanic System and the strong emission features at
$(\ell,b)=(297^\circ,-24^\circ)$ and $(\ell,b)=(302^\circ,-16^\circ)$, is not
visible in the earlier data of Mathewson \& Ford (1984) (their figure
2) or Morras (1982), due to their sparse spatial sampling. 
The total HI mass of the Arm is $\sim 1\times 10^{7}$ M$_\odot$.
 See Putman et~al. (1998) for further details.}
\end{figure}

\begin{figure}
\caption{Peak intensity map which includes regions from Figures 1 and 3, and
velocities $v_{lsr}=165 - 400$ km s$^{-1}$.}
\end{figure}

\begin{figure}
\caption{Channel maps showing the HVC designated \#350 in the Wakker \& van
Woerden (1991) compilation (particularly evident by the knot of emission in the
$v_{lsr}=86$ km s$^{-1}$ channel ($v_{lsr}$ is labelled in the upper left of each panel.)). This is clearly not an isolated
HVC, per se, but one whose emission merges with the neighbouring
Galactic emission.}
\end{figure}

\end{document}